\documentclass[prl,twocolumn,showpacs,amsmath,amssymb]{revtex4}

\usepackage{color}

\usepackage{bm}
\usepackage{dcolumn}% Align table columns on decimal point
\usepackage[dvipdfm]{graphicx}

\begin{document}

\preprint{preprint}

\title{Perfectly conducting channel on the dark surface of weak topological insulators}
\author{Yukinori Yoshimura$^1$}
\author{Akihiko Matsumoto$^1$}
\author{Yositake Takane$^1$}
\author{Ken-Ichiro Imura$^1$}
\affiliation{$^1$Department of Quantum Matter, AdSM, Hiroshima University, Higashi-Hiroshima 739-8530, Japan}

\date{\today}

%%% abstract %%%
\begin{abstract}
A weak topological insulator (WTI) bears, generally, an even number of Dirac cones on its surface;
they are susceptible of doubling, while on the surface of a certain orientation 
it shows no Dirac cone.
On this ``dark'' surface of a WTI, we predict the existence of
a single pair of isolated 1D perfectly conducting channels
that forms either a closed loop or a segment of a line.
The former is associated
typically with a single atomic-layer-thick island
formed on the dark surface, while the latter
is shown to be the consequence
of a pair of crystal (screw) dislocations terminating on the dark surface.
\end{abstract}


\maketitle

The specificity of the three-dimensional (3D)
topological insulator is that it can be weak
\cite{FuKaneMele, MooreBalents, Roy}.
Initially, this weak topological phase has not been paid much attention 
under the pretext that it is not more than a mere
adiabatic deformation of an ordinary insulator.
Yet, it has later turned out that 
in this so-called weak phase
a topological non-triviality 
is not actually weak but rather hidden
\cite{disloc, Stern}.
In contrast to its counter part, the strong topological insulator (STI),
which exhibits a single protected Dirac cone,
the weak topological insulator (WTI) bears either an even number
or, depending on the way the crystal is cleaved, no gapless Dirac cone
on its surface.
And for this very reason, the WTI is often mistaken for possessing
no robust topological character.
%But the point to be made here is that what is said above is indeed the case {\it only with an isolated infinitely large flat surface of a perfect bulk crystal}.

The topological non-triviality ensuring the emergence of
gapless Dirac cones on the surface of topological insulators
is encoded in its 
non-trivial or inverted band structure,
that is to say, stems from a non-trivial feature in the reciprocal space.
In a WTI at least
one of the three so-called 
weak $\mathbb{Z}_2$ indices $\nu_1$, $\nu_2$, $\nu_3$
\cite{FuKane}
remains non-zero,
reflecting its non-trivial band structure.
The status of the STI and WTI could be inverted
when this reciprocal-space feature is interconnected with a topologically 
non-trivial real-space character such as a dislocation of the crystalline structure
in the real space.
In the examples shown in Refs. \cite{Ran_nphys, disloc}
a WTI exhibits a protected gapless state
bound to such a dislocation, 
%\textcolor{blue}{
while this is not necessarily the case with STIs.

The real-space geometry, or the shape of the sample
plays also an important role in the spectrum of the surface state
\cite{prism}.
Indeed, except for the case of an isolated infinitely large single surface,
the surface state is not actually gapless in the strict sense;
the gaplessness is not immune to finite-size gap opening.	 
And the magnitude of this finite-size energy gap is qualitatively 
different in different geometries.
In the simplest and most commonly used geometry of
a slab
%\footnote{slab: a real-space geometry with a doubly periodic condition on the (hypothetical) side surfaces; isomorphic to a torus of a finite thickness.}
such a finite-size energy gap is exponentially small,
and may practically be negligible except in extremely thin film samples
\cite{thin_film}.
In prism-shaped (isomorphic to a cylinder, or a filled torus)
\cite{Vishwanath, Bardarson, aniso}
or cubic (spherical)  
\cite{spherical}
STI,
the magnitude of the size gap is much enhanced,
decaying only algebraically as a function of the linear dimension
of the system.
%since in such geometries the two Dirac cones on the top and the bottom surfaces of the corresponding ``hypothetical slab'' are coupled to each other via the side surfaces.

%%%%%%%%%%%%%%%%%%%
\begin{figure}[t]
\begin{tabular}{c}
\includegraphics[width=60mm]{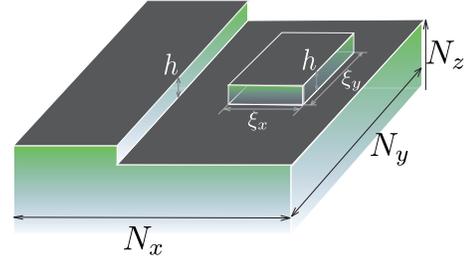}
\end{tabular}
\vspace{-2mm}
\caption{(Color online) Upper and lower terraces separated by a step of height $h$
are examples of the dark surface of WTI. %($\bm\nu$ is pointed normally to these planes). 
On the lower terrace is an island formed with the same height.
%The bulk WTI occupies the lower half of the space.
%Indicated flux insertion is for a thought experiment testing the robustness of the 1D modes formed at the periphery of the island (see main text).
}
\label{schema_terrace}
\end{figure}
%%%%%%%%%%%%%%%%%%%

In contrast to a STI,
a WTI can have ``dark'' surfaces on which no gapless Dirac state appears.
Such dark surfaces can be made
by cleaving the bulk crystal in the direction specified by
the Miller indices $(\nu_1 \nu_2 \nu_3)$,
or equivalently,
by placing the surface normally to the ``weak vector''
$\bm\nu=(\nu_1, \nu_2, \nu_3)$.
As a consequence of the existence of these dark surfaces
WTI samples can exhibit size effects
different from that of a STI of the same shape
\cite{prism}.
Typically, the size effects show {\it even/odd} 
feature with respect to the number of
layers stacking in the direction of $\bm\nu$.
It has also been pointed out that a WTI is not that weak either against disorder
\cite{Stern, Mong, DWTI}.
Here, in this Letter we highlight another striking manifestation of
a hidden non-trivial character of the WTI phase.
The work is indeed motivated by an experimental discovery of such a WTI
and the possibility of cleavage in the plane normal to $\bm\nu$,
{\it i.e.},
realizing a dark surface of WTI
\cite{Kanou}.

%In the first half 
We start by demonstrating that 
a closed loop of perfectly conducting channel (PCC)
is formed at the periphery of
an atomic-layer-thick terrace
%\textcolor{blue}{
of height corresponding to an odd number of atomic layers
formed on a dark surface of WTI.
Experimental observation of 
such atomic-scale terraces 
formed on an otherwise flat surface of a topological insulator
by STM measurements
have been also reported previously
\cite{Yazdani}.

%%%%%%%%%%%%%%%%%%%
\begin{figure}[t]
\includegraphics[width=70mm]{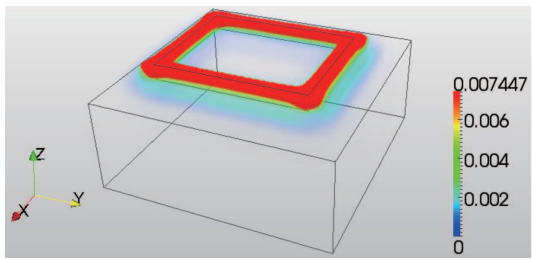}
\vspace{-2mm}
\caption{(Color online) A pseudo-gapless 1D mode emergent
along the periphery of an island. 
The island has a size of $\xi_x \times \xi_y = 9 \times 11$, 
and an altitude $h =1$.
The systems's sizes are $N_x=N_y=16, N_z=7$.
}
\label{wf_island}
\end{figure}
%%%%%%%%%%%%%%%%%%%

Let us consider the following standard Wilson-Dirac type
effective Hamiltonian for 3D topological insulators,
here
implemented on the cubic lattice as
\begin{eqnarray}
H_{\rm bulk} &=& A \tau_x \sigma_\mu \sin k_\mu +\tau_z m(\bm k), 
\nonumber \\
m(\bm k) &=& m_0 + 2 m_{2\mu} (1-\cos k_\mu),
\end{eqnarray}
where two types of Pauli matrices $\bm \sigma$ and $\bm \tau$
represent real and orbital spins, and $\mu=x, y, z$.
The parameter $A$ represents the strength of normal hopping, and
determines the aperture of the surface Dirac cone.
To realize a WTI with a specific choice of weak indices
$\bm\nu=(0, 0, 1)$,
we choose the mass parameters to be anisotropic.
In the following demonstrations they are chosen such that
$m_0=-1$, $m_{2x}=m_{2y}=0.5$ and $m_{2z}= 0.1$
in units of the hopping parameter $A$ \cite{prism}.
In the actual simulations on-site random potential is also introduced 
as in Ref. \cite{DWTI}.
The global shape of the sample is also assumed to be tetragonal
apart from terraces and islands; see below,
{i.e.},
the underlying lattice is,
unless otherwise mentioned,
terminated in all the three ($x$-, $y$- and $z$-)
directions, and in principle could accommodate a surface state
on all of its tetragonal faces, but because of the weak vector 
pointing to the $\hat{\bm z}$-direction
that excludes Dirac cones from the surfaces normal to this direction,
the system exhibits pseudo-gapless
%\footnote{The surface states are slightly gapped by finite-size effects of two different origins: (i) finite-size effects due to the half-odd integral quantization of ``$L_z$'' in the language of (topologically equivalent) cylindrical system with a full rotational symmetry around the $z$ axis. (ii) parity of $N_z$ dependent effect, specific to WTI.\cite{prism} See also the main text.}
surface states only 
on the four side surfaces.

Let us now consider a situation in which
the top surface is not completely flat, 
showing either atomic-scale terrace-like step structures
or an isolated island-shaped region
as depicted in Fig. \ref{schema_terrace}.
%%%%%%%%%%%%%%%%%%%%%%%
%\textcolor{blue}{
When the surface is normal to $\bm \nu$, the vertical wall of this step structure
is parallel to $\bm \nu$, 
on which 
one can hypothesize formation of gapless 2D surface states (two Dirac cones).
Then, of course,
contributions from these two Dirac cones
must be superposed to cope with the boundary condition
that the wave function is not allowed to
%\textcolor{blue}{
leak
onto the dark terraces \cite{prism}.
%%%%%%%%%%%%%%%%%%%

\begin{figure}[t]
(a)
\includegraphics[width=70mm]{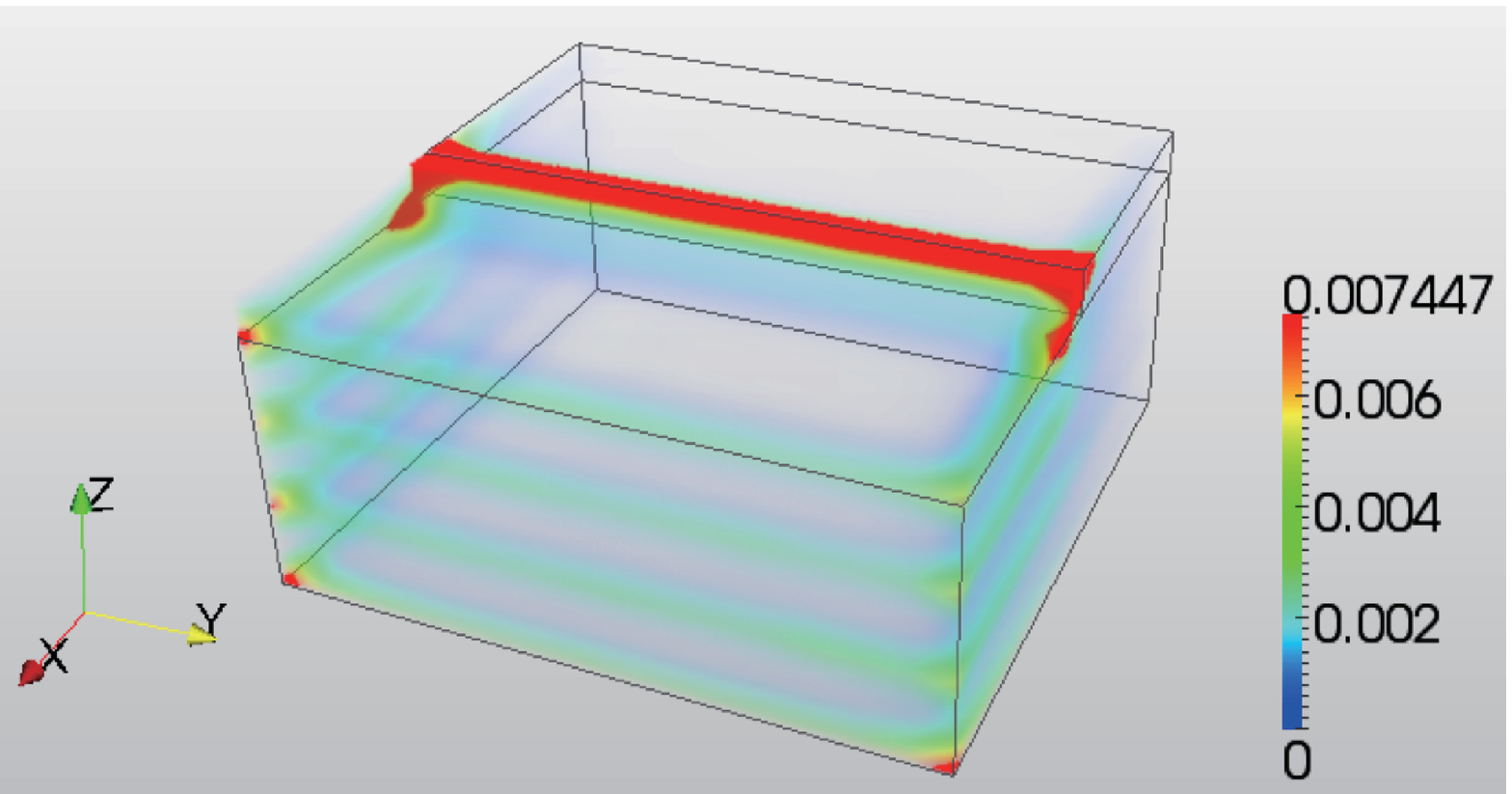}
\\
(b)
\includegraphics[width=70mm]{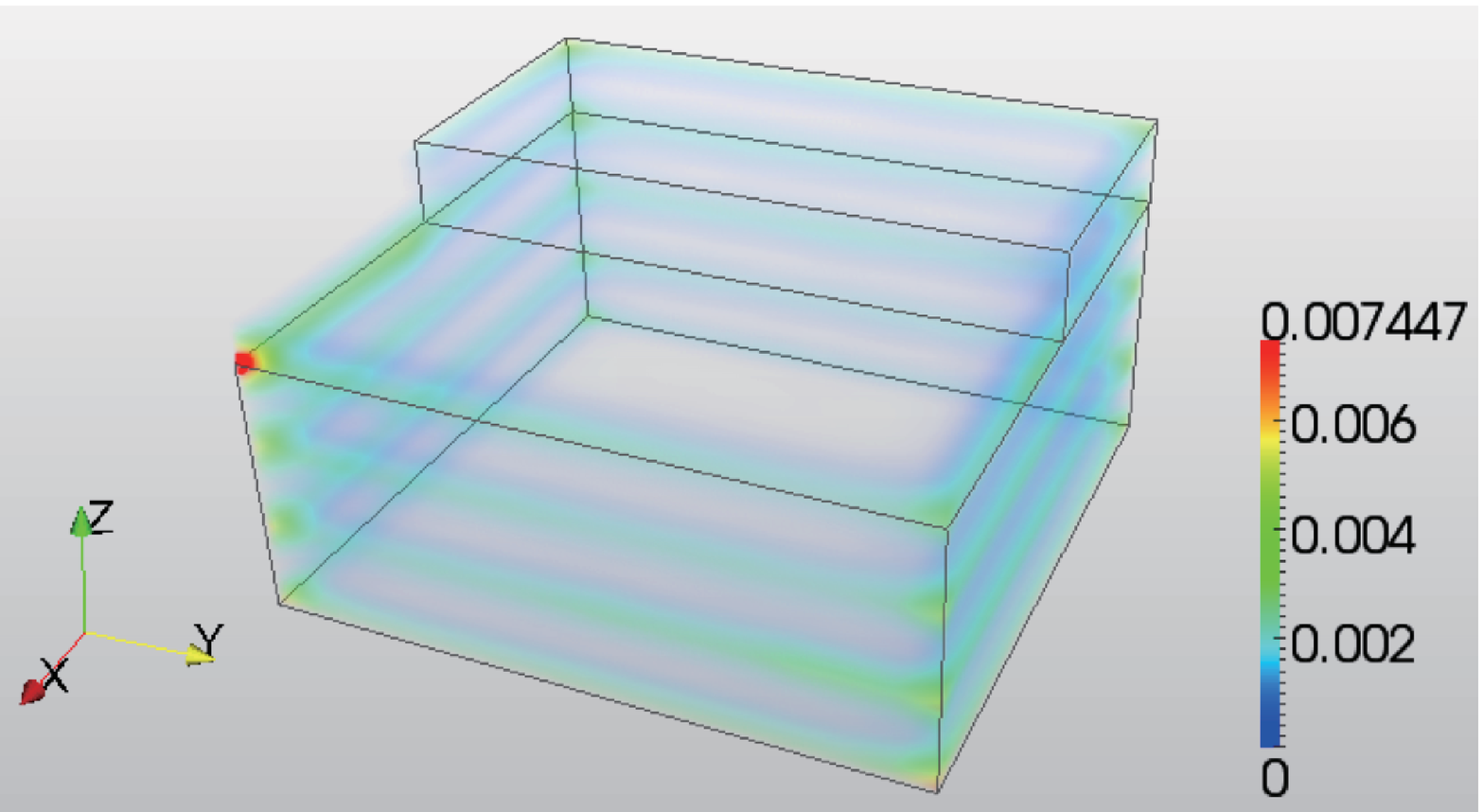}
\vspace{-2mm}
\caption{(Color online) Profile of the lowest energy wave function in
atomic-scale terraces of a WTI. %as depicted in Fig. 1.
(a) When $h=1$, 1D modes appear at the edge of the terrace.
(b) when $h=2$,
the wave function has a vanishing amplitude
along the edge of the terraces.
$N_x=N_y=16, N_z=7$.
}
\label{wf_terrace}
\end{figure}

Fig. \ref{wf_island} and
Fig. \ref{wf_terrace} (a) 
demonstrate that there appears indeed
a pair of 1D modes almost localized
at the periphery of such a terrace- or an island-shaped region,
when its height $h$ (measured in units of the lattice constant) 
is odd ($h=1,3,5,\cdots$).
These 1D modes
%enclosing the square-shaped island 
are,
in contrast to 2D surface states,
%\footnote{Notice that in WTI's the surface states get affected by the ``inter-valley'' scattering between the two Dirac cones.}
immune to backward scattering as a consequence of
their helical nature,
and realize an example of what is often called ``perfectly conducting channels''
\cite{pcc}.
The wave functions associated with these 1D channels
shown in Fig. \ref{wf_island} and Fig. \ref{wf_terrace} (a) 
are barely influenced by
the presence of weak disorder;
here, the strength of disorder is chosen as
$W/A=1$ in the notation of \cite{prism}.
%while those of 2D surface state tend to get localized [see Fig. \ref{wf_terrace} (b)].
%\textcolor{blue}{
The above numerical simulations have been performed in systems of
a rather simple shape of the island and of the terrace.
While, the appearance of a 1D PCC is considered to be generic, 
and equally applicable to systems that are smoothly connected to the present one
as far as $h$ is odd.

%%%%%%%%%%%%%%%%%%%
%%%%%%%%%%%%%%%%%%%

If $h$ is even, the situation is much different as a result of the {\it even/odd} feature 
of finite-size gap generation
characteristic to the surface of WTI
\cite{prism}.
This {\it even/odd} feature is associated with the fact that
a WTI is sometimes described 
as stacked layers of 2D quantum spin Hall states,
each exhibiting a pair of helical edge modes.
These helical edge modes, when stacked to form a helical surface state,
tend to couple into gapped pairs, 
leaving only a single gapless channel when the number %$N_z$ 
of stacked layers is odd.
A more quantitative $\bm k\cdot\bm p$-type argument 
developed in Ref. \cite{prism}
suggests when applied to the terrace- or island-shaped region 
%formed on top of the cleaved surface in the $\hat{\bm z}$-direction,
that the pseudo-zero-energy states acquire a finite energy,
$E_\pm=\pm {A\over 2(h+1)}\pi$,
%\textcolor{blue}
when its height $h$ is even.
%where the parameter $A>0$ determines the slope of helical Dirac cone.
%\textcolor{blue}{
%Notice that the consequence of this finite-size effect is incomparable to the case of the same effect on the 2D surface spectrum  \cite{prism}. An analogous finite-size correction to the 2D surface state results in a much smaller effect that might be eventually washed out by other effects such as those of disorder.
%%%%%%%%%%%%%%%%%%
Here, the pseudo-gapless 1D helical modes are completely kicked out of
the low-energy spectrum
[see Fig. \ref{wf_terrace} (b)].
The case of a two-story ($h=2$) terrace is an example in which 
this %parity of the layer number
%\textcolor{blue}{
(parity of) $h$-dependent
size effect is most
accentuated, giving rise to a huge finite-size correction 
to the energy of pseudo-zero modes as, $E_\pm=\pm {A\over 6}\pi$,
{\it i.e.},
on the order of 1 in units of $A$.

The protected 1D modes emergent
when $h$ is odd, typically one,
are not entirely immune to finite-size gap opening, %either
but of a different, less destructive type.
On one hand,
these counter propagating modes are always in the
opposite spin state (spin-to-momentum locking), and
this very fact makes them immune to
backward scattering by the impurities.
While, on the other hand,
the quantization axis of these spins 
can and does indeed rotate
as an electron propagates along this circular 1D orbital,
in such a way that this quantization axis is normal
to the tangent of this 1D orbital (spin-to-surface locking);
a situation analogous to the one occurring at a cylindrical surface of STI
\cite{Vishwanath, Bardarson, aniso}.
As a result, while an electron revolves once around a circular orbit,
its spin also performs a complete $2\pi$ rotation, altering the orbital part of the
boundary condition from periodic to anti-periodic.
This leads,  tracing exactly the same path as in the case of the half-odd integral quantization
of the orbital angular momentum $L_z$ in a cylindrical STI
\cite{aniso,prism},
to a finite-size correction $E_\pm=\pm {\tilde{A}\over s}\pi$,
where $\tilde{A}$ is a constant of order $A$, and
$s=4\xi$ is the circumference of the island.
%the magnitude of which is still much smaller than that of Eq. (\ref{Nh}).

%%%%%%%%%%%%%%%%%%%
\begin{figure}[t]
\includegraphics[width=70mm]{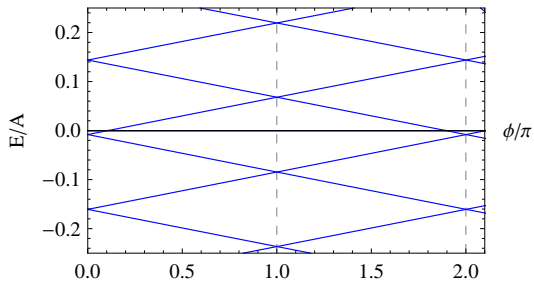}
\vspace{-2mm}
\caption{(Color online) A numerical experiment: 
verification of the perfectly conducting character of the 1D channels
circulating the periphery of the island
(case of $h =1$).
The spectrum of the 1D modes is plotted against the 
inserted flux $\phi$.}
\label{spec_phi}
\end{figure}
%%%%%%%%%%%%%%%%%%%

To double check 
that the pseudo-gapless 1D modes in the case of $h$ being odd
realize indeed a single pair of perfectly conducting channels,
and that they are immune to backward scattering by the impurity,
we have performed the following numerical experiment.
In a situation as depicted in Fig. 2,
in which
a 1D ring-shaped distribution of the wave function is realized along the
periphery of an island,
we gradually introduce a magnetic flux tube (of magnitude $\Phi$)
threading the island
%\textcolor{blue}{
such that ideally the flux would not touch the electron.
The boundary condition on the side surfaces are here chosen to be
doubly periodic, {\it i.e.}, a slab geometry is employed
to avoid 2D states on the side surfaces.
%\footnote{Of course, in numerical experiments in a finite-size system this can be achieved only up to the accuracy with which the wave function of the 1D modes decay exponentially toward the center of the island.}
%\textcolor{blue}{
Fig.~\ref{spec_phi} depicts the result of such a numerical experiment
in which the spectrum of the 1D states
is plotted against $\phi$.
Here,
$\phi = 2\pi {\Phi\over \Phi_0}$ is an Aharonov-Bohm phase associated with
the external flux $\Phi$ measured in units of the flux quantum $\Phi_0$.
One can clearly observe that 
(i) the spectrum is doubly degenerate at $\phi=0$ and at $\phi=\pi$,
(ii) as $\phi$ is introduced, say at $\phi=0$,
this two-fold degeneracy is lifted in such a way that
one state is shifted upward, and the other downward,
indicating the existence of counter propagating channels.
Note also that
(iii) at $\phi=0$ and at $\phi=\pi$, the pairs are swapped, and
(iv) the $\phi$-dependence of each spectrum is almost perfectly linear.
These features are %as we further argue below
perfectly in accordance with our hypothesis that
a pair of 1D helical modes are formed at the periphery of the island.
%\textcolor{blue}{Note also that such a regular pattern associated with the 1D modes exhibits a sharp contrast to non-robust behaviors that can be also seen around $\phi\sim 4\pi$ in Fig.~\ref{spec_phi} due to the formation of bound states associated with the flux, here inserted as four separated pieces: $\Phi \rightarrow \Phi/4 \times 4$ to make such unwanted bound states clearly distinguishable from the signatures of the 1D PCC.}

%\textcolor{blue}{
The spectrum of the 1D helical modes in the presence of a flux $\Phi$
can be written most generally as
\begin{equation}
E^{(\pm)}_n (\phi) = \pm \tilde{A} {\pi \over s}
\left(n-{1\over 2}-{\phi\over2\pi}\right) +\Delta E,
\label{spec_1D}
\end{equation}
where $\Delta E$ represents a global shift of the spectrum
associated with the breaking of the particle-hole symmetry
due to on-site random potentials.
%and $L=4 N_{isl}$ is the circumference of the island. 
At $\phi=0$,
the two-fold degeneracy of the spectrum occurs such that
$E^{(+)}_n (0) = E^{(-)}_{-n+1} (0)$,
while at $\phi=\pi$, after these partners are shifted in the opposite directions,
perfectly linearly to $\phi$,
each state gets paired again but with a different partner
such that
$E^{(+)}_n (\pi) = E^{(-)}_{-n} (\pi)$,
{\it i.e.}, in such a way that
the relative ordering of the partners is shifted by one
\cite{NKR}.
%The immunity of the crossings at these points against backward scattering by the impurities ensures that these 1D modes realize indeed a PCC.
%\footnote{L. D. Landau and L. M. Lifshitz, Quantum mechanics: the non-relativistic theory, 3rd ed., (Butterworth-Heinemann, Stoneham, MA, 1981), Sec. 79.}.

%%%%%%%%%%%%%%%%%%%
\begin{figure}[t]
\includegraphics[width=70mm]{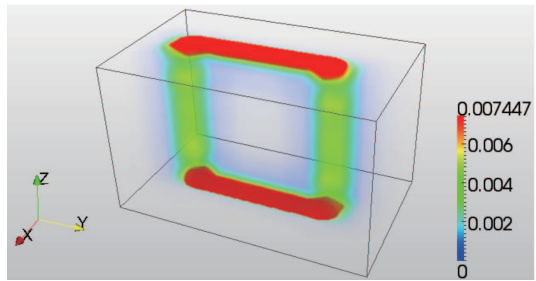}
\vspace{-2mm}
\caption{(Color online) A finite segment of 1D gapless modes
penetrating into the bulk WTI at the ends of the segment
along a screw dislocation line 
with $\bm b =b\hat{\bm z}$ $(b=\pm 1)$.
The two dislocation lines are separated by a distance of
$\zeta=8$.
$N_x=16, N_y=N_z=10$.
}
\label{wf_disloc}
\end{figure}
%%%%%%%%%%%%%%%%%%%

We have seen so far
examples in which pseudo-gapless protected 1D modes
appear on a dark surface of WTI,
forming
either a closed loop, {e.g.}, at the periphery of an island,
or a line, {e.g.}, at the boundary of two terraces, crossing the dark surface
completely,
only to merge with a 2D side-surface state at both ends. 
In the remainder of this Letter
we focus on another type of 1D modes
emergent on the dark surface,
typically as {\it a finite segment of a line}
of length $\zeta$.
We demonstrate that this type of 1D modes are %necessarily 
associated with
the presence of dislocation lines in the crystal,
and indeed the two ends of the segment
coincide with the termination of a dislocation line at the dark surface
(see Fig. \ref{wf_disloc}).

To convince oneself that this is certainly the case
let us first turn around the argument in the following way.
Recall that a pair of gapless 1D helical modes
are induced along a dislocation line introduced
in the bulk WTI
\cite{Ran_nphys, disloc},
when the Burgers vector $\bm b$ specifying the dislocation
satisfies the relation,
$\bm b\cdot\bm M = \pi \mod 2\pi$,
where the vector $\bm M$ is defined in terms of the weak indices 
and the three reciprocal lattice vectors $\bm G_1$, $\bm G_2$, $\bm G_3$
%associated with these indices 
as $\bm M={1\over 2}(\nu_1 \bm G_1 + \nu_2 \bm G_2 + \nu_3 \bm G_3)$.
The question to be addressed here can then be reformulated in the following way.
What happens to
the single pair of gapless 1D helical modes
that emerges along a dislocation line,
when the dislocation is terminated at a dark surface of WTI?
In any realistic setup
it is natural to assume that a dislocation line is terminated at
the surface of the sample at both ends,
unless it forms a closed loop.
Here, we consider a dislocation line 
%that does not form a closed loop, but are 
terminated by the two {\it dark} surfaces, 
such as the ones on the top and on the bottom
%\textcolor{blue}{
(see Fig. \ref{wf_disloc}).
Since
these dark surfaces are normal to $\bm \nu$,
the condition for the existence of gapless 1D helical modes %(mentioned previously)
%simultaneously 
is satisfied, 
only when the introduced dislocation is a {\it screw} dislocation
with a dislocation line %$\bm l$ 
parallel to $\bm b$.
When such a screw dislocation 
is terminated at the surface of a sample,
it necessarily accompanies 
a mismatch of atomic layers by a magnitude of $b=|\bm b|$,
forming typically a step structure,
analogous to the one that appears between the two terraces.

%In the original experimental point of view, what could be observed is 
%\textcolor{blue}{
Let us come back to the original point of view 
in which our focus is on
a finite segment of the 1D helical modes emergent on the dark surface.
Notice that
a single pair of such Dirac states
cannot be confined, in principle,
in a finite segment
(Klein tunneling).
The only way (out) for such a segment of 1D modes 
to appear on the dark surface is
that it pair-annihilates with another 1D helical modes
along a dislocation line that terminates at the dark surface.
This is explicitly demonstrated in Fig. \ref{wf_disloc}.

%%%%%%%%%%%%%%%%%%%
\begin{figure}[t]
\includegraphics[width=70mm]{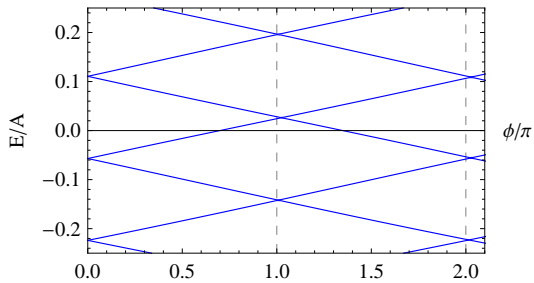}
\vspace{-2mm}
\caption{(Color online) The ``flux experiment'' applied to a configuration 
with a pair of screw dislocation lines.}
\label{spec_phi_disloc}
\end{figure}
%%%%%%%%%%%%%%%%%%%

The last question to be addressed is whether these
continuous 1D channels partly hosted by a dislocation line,
combined with a segment of 1D modes on the dark surface
realize also a perfectly conducting channel.
For this purpose
we have performed a numerical
experiment,
analogous to the one for the island geometry, 
but here, %as shown in Fig. \ref{schema_disloc},
the flux tube is inserted between the two dislocation lines.
In Fig. \ref{spec_phi_disloc} 
one can clearly recognize
the four characteristic features
supporting the formation of a pair of 1D helical modes
observed in the case of the island.
%The ``avoided level crossing'' at $\phi=0$ and at $\phi=\pi$ shown in 
Here, in the present geometry this
confirms that
different parts of the helical modes,
one on the dark surface,
and the other along a dislocation line,
are indeed smoothly connected
to form a 1D perfectly conducting channel.
Here, $s$ in Eq. (\ref{spec_1D}) is given by
$s=2(\zeta+N_z)$, the circumference of the 1D channel.

A single pair of 
pseudo-gapless 1D helical modes also appears
along a dislocation line inserted in a STI.
However, when the pair is hosted by a STI, 
it necessarily terminates in the ``sea'' of the single-cone Dirac state
at the surface where surface sensitive measurements
are applicable.
Here, we have provided a more experimentally feasible situation
to detect a dislocation solely from information on the surface,
by predicting that {\it termination} of a 1D perfectly conducting channel
on the dark surface of WTI %necessarily 
accompanies a screw dislocation line 
in the bulk host crystal
that terminates at the dark surface.

KI acknowledges M. Kanou and T. Sasagawa for stimulating discussions.
KI and YT are supported by KAKENHI; 
KI by the ``Topological Quantum Phenomena'' (No. 23103511), 
and YT by a Grant-in-Aid for Scientific Research (C) (No. 24540375).

\end{document}